\documentclass[usegraphicx,usenatbib,useAMS]{mn2e}

\usepackage{times} 
\usepackage{aas_macros}     
\usepackage{amssymb}        

\newcommand{\bc}{\begin{center}}
\newcommand{\ec}{\end{center}}

\newcommand{\lsim}{\lesssim} 

\title{ Which halos host Herschel-ATLAS galaxies in the local Universe?}
\author[Qi Guo et al.]
    {\parbox{18cm} { \large Qi Guo$^1$, Shaun Cole$^1$,  Cedric
G. Lacey$^1$, Carlton M. Baugh$^1$, Carlos S. Frenk$^1$, Peder
Norberg$^2$, R. Auld$^4$, I. K. Baldry$^{10}$, S. P. Bamford$^{3}$,
N. Bourne$^3$, E. S. Buttiglione$^7$, A. Cava$^8$,
A. Cooray$^{17}$, S. Croom$^{15}$, A. Dariush$^4$, G. De Zotti$^{7,20}$,
S. Driver$^{11}$, L. Dunne$^3$, S. Dye $^4$ S. Eales$^4$, J. Fritz$^9$,
A. Hopkins$^{12}$, R. Hopwood$^{19}$, E. Ibar$^5$,
R. J. Ivison$^5$, M. Jarvis$^{18}$, D. H. Jones$^{12}$,
L. Kelvin$^{11}$, J. Liske$^{13}$, J. Loveday$^{14}$,
S. J. Maddox$^3$, H. Parkinson$^{2}$, E. Pascale$^4$,
J. A. Peacock$^{2}$,  M. Pohlen$^4$, M. Prescott$^{16}$,
E. E. Rigby$^3$, A. Robotham$^{11}$, G. Rodighiero$^7$,
R. Sharp$^{12}$, D. J. B. Smith$^4$, P. Temi$^6$, E. van Kampen$^{13}$ } 
       \\
       \\
       $^1$ Institute for Computational Cosmology, Department of Physics, University of Durham, South Road, Durham, DH1 3LE, UK \\ 
       $^2$ SUPA, Institute for Astronomy, University of Edinburgh, Royal Observatory, Blackford Hill, Edinburgh, EH9 3HJ, UK \\   
       $^3$ School of Physics and Astronomy, University of Nottingham,
       Nottingham, NG7 2RD, UK \\
       $^4$ School of Physics \& Astronomy, Cardiff University, CF24
       3AA, UK \\
       $^5$ UK Astronomy Technology Centre, Royal Observatory, Edinburgh, EH9 3HJ, UK \\
       $^6$ Astrophysics Branch, NASA Ames Research Center, Mail Stop 2456, Moffett Field, CA 94035, USA\\
       $^7$ INAF ­ Osservertorio Astronomico di Padova, Vicolo Osservatorio 5, I-35122, Padova, Italy \\
       $^8$ Instituto de Astrof´sica de Canarias (IAC) and Departamento de Astrofisica de La Laguna (ULL), La Laguna, Tenerife, Spain\\
       $^9$ Sterrenkundig Observatorium, Universiteit Gent, Krijgslaan 281 S9, B-9000 Gent, Belgium\\
       $^{10}$ Astrophysics Research Institute, Liverpool John Moores University, Twelve Quays House, Egerton Wharf, Birkenhead CH41 1LD\\
       $^{11}$ School of Physics and Astronomy, University of St
       Andrews, North Haugh, St Andrews, Fife, KY16 9SS \\
       $^{12}$  Australian Astronomical Observatory, PO Box 296, Epping, NSW 1710, Australia \\
       $^{13}$  European Southern Observatory, Karl-Schwarzschild-Strasse 2 D-85748, Garching bei Munchen, Germany\\
       $^{14}$ Astronomy Centre, University of Sussex, Falmer,
       Brighton, BN1 9QH, UK \\
      $^{15}$ Sydney Institute for Astronomy, School of Physics, University of Sydney, NSW 2006, Australia \\
   $^{16}$  Astrophysics Research Inst., Liverpool John Moores University, 12 Quays House, Egerton Wharf, Birkenhead, CH41 1LD, UK\\
   $^{17}$ Department of Physics and Astronomy, University of California, Irvine, CA 92697, USA \\
    $^{18}$ Centre for Astrophysics, Science \& Technology Research Institute, University of Hertfordshire, Hatfield, Herts, AL10 9AB\\
    $^{19}$ Department of Physics and Astronomy, The Open University,
    Walton Hall, Milton Keynes MK7 6AA, UK\\
    $^{20}$ SISSA, Via Bonomea 265, I-34136 Trieste, Italy\\
    } 
       
\begin{document}

\maketitle

\begin{abstract}
We measure the projected cross-correlation between low redshift ($z <
0.5$) far-IR selected galaxies in the SDP field of the {\it
Herschel}-ATLAS (H-ATLAS) survey and optically selected galaxies from
the Galaxy and Mass Assembly (GAMA) redshift survey. In order to
obtain robust correlation functions, we restrict the analysis to a
subset of 969 out of 6900 H-ATLAS galaxies, which have reliable
optical counterparts with $\rm r<19.4$\,mag and well-determined
spectroscopic redshifts. The overlap region between the two surveys is
12.6~deg$^2$; the matched sample has a median redshift of $z\approx
0.2$.  The cross-correlation of GAMA and H-ATLAS galaxies within this
region can be fitted by a power law, with correlation length
{$r_0\approx 4.63 \pm 0.51$~Mpc}.  Comparing with the corresponding
auto-correlation function of GAMA galaxies within the SDP field yields
a relative bias (averaged over 2-8\,~Mpc) of H-ATLAS and GAMA galaxies
of $b_{\rm H}/b_{\rm G} \approx 0.6$. Combined with clustering
measurements from previous optical studies, this indicates that most
of the low redshift H-ATLAS sources are hosted by halos with masses
comparable to that of the Milky Way.  The correlation function appears
to depend on the 250~$\mu$m luminosity, $\rm L_{250}$, with bright
(median luminosity $\nu \rm L_{250}
\sim 1.6 \times 10^{10}$L$_{\odot}$) objects being somewhat more strongly
clustered than faint ($\nu \rm L_{250} \sim 4.0 \times 10^{9}$L$_{\odot}$)
objects. This implies that galaxies with higher dust-obscured star
formation rates are hosted by more massive halos.

\end{abstract}

\section{Introduction}

It is well known that $L^*$ galaxies are the largest contributors to
the present-day stellar mass density \citep[e.g.][]{Li2009}. It is,
however, not clear how star formation is distributed across galaxies
and halos of different masses. Previous studies show that in the local
universe star formation takes place preferentially in low density
environments \citep[e.g.][]{Lewis2002,Heinis2009}.  The most commonly
used estimators of the star formation rate (SFR) are based on the UV
continuum or H$\alpha$, H$\beta$ or [OII] emission lines
\citep[e.g.][]{Brinchmann2004,Salim2007}.  These are all subject to
uncertain dust extinction corrections, and so can greatly
underestimate the SFRs in dust-obscured regions. Mid- and far-IR
observations, which are sensitive to the energy re-emitted by dust
heated by young stars, are therefore an essential complement to UV and
optical tracers of star formation.  Such dust is heated to
temperatures of around $20-40$~K, emitting thermal radiation which
peaks at wavelengths around $100$~$\mu$m. IRAS measured the far-IR
emission from bright galaxies but more recent surveys of dust emission
have focussed on either mid-IR (ISO, Spitzer) or sub-mm (e.g. SCUBA)
wavelengths, missing the peak in the dust emission, and therefore
requiring uncertain extrapolations to infer total IR luminosities and
hence dust-obscured SFRs. The launch of Herschel
\citep{Pilbratt2010} has now opened up the study of the universe at
far-IR wavelengths ($60-700~\mu$m), spanning the peak of the dust
emission from star-forming galaxies, and allowing robust measurements
of the dust-obscured SFR. The Herschel-ATLAS (H-ATLAS) survey
\citep{eales2010} will provide far-IR imaging and photometry covering
the wavelength range from 110~$\mu$m to 500~$\mu$m, over an area 550
deg$^2$, much larger than previous surveys at these wavelengths such
as BLAST \citep{Devlin2009}.

Analysis of clustering statistics provides a simple but powerful way
to investigate environmental effects, in this case of the SFR of
galaxies. In this paper we perform a preliminary clustering analysis
of a 4$\times$4 deg$^2$ field observed during the H-ATLAS science
demonstration phase (SDP). Previous analyses of the H-ATLAS
\citep{Maddox2010} and HerMES \citep{Cooray2010} surveys have focused
on angular auto-correlations, with no significant signal in the former
case and a significant detection in the latter. Here, we consider
spatial cross-correlations of far-IR and optical galaxies, which can
be used to derive the clustering bias and hence the characteristic
mass of the host halos. We analyze a sample of $\sim$1000 H-ATLAS
galaxies which have reliable counterparts brighter than ${\rm r <
  19.4}$\,mag in the SDSS and spectroscopic redshifts measured by the
Galaxy and Mass Assembly (GAMA) survey\footnote{GAMA will eventually
  provide a highly complete, wide-area spectroscopic survey of over
  400,000 galaxies with sub-arcsecond optical/near-IR imaging (from
  SDSS, UKIDSS, VST, VISTA), and complementary observations from the
  UV (GALEX) through to the mid and far-IR (WISE, HERSCHEL) and the
  radio (ASKAP, GMRT). GAMA has so far surveyed 144 deg$^2$ and the
  catalogue contains 95,000 galaxy redshifts to r-band magnitude 19.4
  with a redshift completeness of 98.7\% \citep{Driver2010,Driver2009,
    Baldry2010, Robotham2010, Hill2010}.}.  Currently, the overlap
region between the H-ATLAS and GAMA surveys is 12.6 deg$^2$, and the
spectroscopic redshift completeness is 99.7\% for galaxies with $\rm
r<19.4$\,mag.

A full analysis of the spatial auto-correlation function of H-ATLAS
galaxies is given in \citet{vanKampen2010}. Here we instead measure
the cross-correlation function of H-ATLAS and GAMA galaxies, a
statistic that provides a more robust and accurate estimate of the
clustering bias of the H-ATLAS galaxies. There are at least two
reasons why this is so.  Firstly, the sample of H-ATLAS in the
relatively small SDP survey area is small. By contrast, the number of
GAMA galaxies in this area exceeds that of H-ATLAS galaxies by a
factor of $\sim$ 10. Secondly, the redshift distribution of the GAMA
galaxies can be robustly measured from the full GAMA survey (rather
than from just the restricted SDP area) and for the estimator we
employ knowledge of the H-ATLAS redshift distribution is not
required. Thus the systematic
uncertainties due to cosmic variance are reduced. 
As a result, the estimate of the
cross-correlation function of the relatively sparse H-ATLAS sample
with the more populous GAMA sample has much better statistics than the
estimate of the H-ATLAS auto-correlation function alone.  Finally,
even though our sample is relatively small, using the
cross-correlation technique allows us to investigate the dependence of
clustering on far-IR luminosity by dividing the H-ATLAS sample into
two subsets according to 250~$\mu$m luminosity.  In this manner, we
determine the clustering bias and infer the typical halo mass for each
subset.

Throughout this paper we assume a flat $\Lambda$CDM cosmology with
$\Omega_{\rm m}$ = 0.25, $\Omega_{\Lambda}$ = 0.75 and H$_0$ =
73~km~s$^{-1}$~Mpc$^{-1}$.

\section{Sample selection}
\label{sec:sample}
We use data obtained by the Spectral and Photometric Imaging Receiver
\citep[SPIRE,][]{Griffin2010,Pascale2010} in the 16~deg$^2$ H-ATLAS
science demonstration field\footnote{PACS data, Ibar et al. 2010, are
also available but are not used here.}. In total there are $6,878$
sources over an area 14.4~deg$^2$ that are brighter than the 5$\sigma$
detection limit in one or more of the 3 SPIRE bands: 250~$\mu$m,
350~$\mu$m and 500~$\mu$m \citep{Rigby2010}. The corresponding flux
limits are 33, 36 and 45~mJy/beam. Below we work with the 250~$\mu$m flux limited sample as this is
the most sensitive band, has the best positional accuracy and was used
for source detection in the catalogue that was matched to GAMA
\citep{Smith2010}.

A significant fraction of these 6,878 Herschel galaxies lie at low
redshifts and have optical counterparts in the SDSS imaging catalogue.
Sources with S/N$\ge$5 at 250~$\mu$m (6,621) were matched to the
r-band selected ($\rm r<$22.4) SDSS catalogue by \cite{Smith2010} using a
likelihood ratio analysis \citep{Sutherland1992, Ciliegi2003} with a
maximum $10^{\prime\prime}$ search radius.
This leads to 4,756
sources which have at least one candidate optical counterpart in SDSS.
A reliability value ($R_{\rm LR}$) is then assigned to each of the
optical candidates, which quantifies the probability that the
counterpart is a genuine match.  We discard candidates with $R_{\rm
LR} < 0.8$ to remove unreliable matches, leaving 2,424 reliably
matched sources. The angular overlap of the GAMA 9-hour field with
H-ATLAS is not perfect and this reduces the survey region (hereafter
GAMA-SDP) from 14.4~deg$^2$ to 12.6~deg$^2$. Within this region there
are 2,143 reliably matched sources. The spectroscopic redshift
coverage of GAMA in this region is complete at 99.7\% for an r-band
Petrosian magnitude (corrected for Galactic extinction) brighter than
19.4\,mag. Imposing this cut leaves 969 galaxies which have measured
spectroscopic redshifts and form the sample we analyse below (the
H-ATLAS sample). A statistical analysis of the excess number of
close pairs shows that 16\% of GAMA sources 
brighter than 19.4\,mag have a Herschel-ATLAS counterpart, of these, 
$\sim$80\% have directly identified reliable matches
\citep{Smith2010}. We only have spectroscopic redshifts for H-ATLAS
galaxies that have reliable matches in the r-band limited GAMA survey.
Hence while we believe we a have complete representative sample of the local
H-ATLAS galaxies we could in principle be missing galaxies which are
bright at 250~$\mu$m but too faint for detection in the r band.  
This possibility can not be ruled out until we have
spectroscopic redshifts selected in the sub-mm.

\begin{figure}
\bc
\hspace{-0.6cm}
\resizebox{8.5cm}{!}{\includegraphics{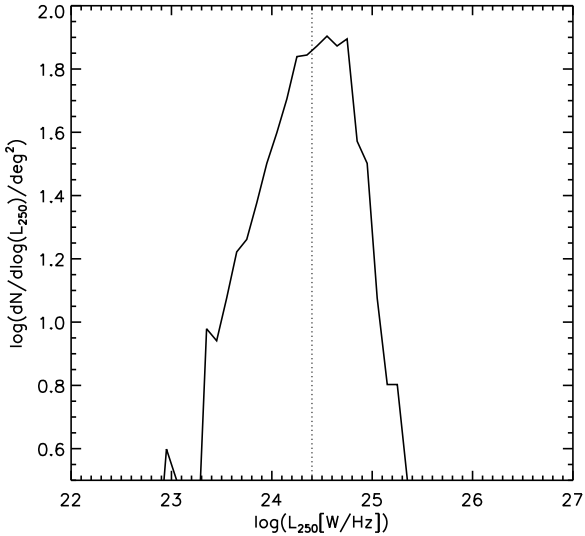}}
\caption{Observed distribution of Herschel 250~$\mu$m rest-frame
 luminosities for the 969 galaxies well matched to SDSS galaxies with
 $\rm r<$19.4\,mag. The dotted line corresponds to the threshold ($2.5 \times 10^{24}$~W~Hz$^{-1}$) used to
 split the sample into bright and faint subsets (see text Section~\ref{sec:sample}).}
\label{fig:Lum}
\ec
\end{figure}

To k-correct the observed Herschel fluxes to the rest-frame
250~$\mu$m, we assume that the dust emission has an SED of the form
\begin{equation}
\rm L_{\nu} \propto B_{\nu}(T)\, {\nu}^{\beta},
\label{eq:Tbeta}
\end{equation}
where $B_{\nu}(T)$ is the Planck function. There are two parameters in
this formula, the dust temperature, $T$, and the emissivity index, 
$\beta$.  We adopt the values, $T = 28$~K and $\beta= 1.5$, derived by
\cite{Amblard2010} by fitting to nearby H-ATLAS galaxies
detected in at least three far-IR bands with a significance greater
than 3$\sigma$.  

\begin{figure}
\bc
\hspace{-0.6cm}
\resizebox{8.5cm}{!}{\includegraphics{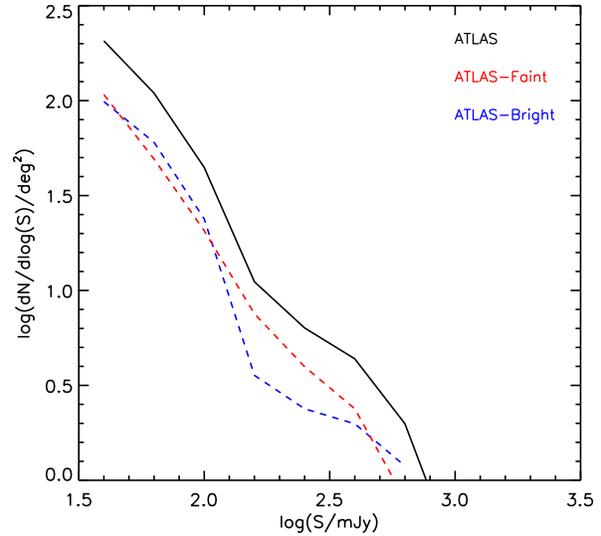}}
\caption{Flux distribution of Herschel sources at 250~$\mu$m. Blue and
red dashed curves are for the bright and faint Herschel sources
respectively, with the black line showing the combination of the two.}
\label{fig:flux250}
\ec
\end{figure}

The luminosity distribution at 250~$\mu$m ($\rm L_{\rm 250}$) is shown
in Fig.~\ref{fig:Lum}. It peaks at around $\rm
L_{250}=3.2 \times 10^{24}$~W~Hz$^{-1}$, corresponding to the local
$L^*$ galaxies found by \cite{Dye2010}.  We further split the H-ATLAS
sample into two subsets (indicated by the vertical dotted line in
Fig.~\ref{fig:Lum}): bright sources with $\rm
L_{250}>2.5 \times 10^{24}$~W~Hz$^{-1}$ (corresponding to total IR
luminosity, $\rm L_{IR}= 5.0 \times 10^{10}$L$_{\odot}$, based on Eqn~1,
integrating from 8 -1000 $\mu$m), and faint sources with $\rm
L_{250}<2.5 \times 10^{24}$~W~Hz$^{-1}$. The faint subset consists
of 484 galaxies and the bright one of 485 galaxies. The median values
of $\rm L_{250}$ for the faint and bright H-ATLAS subsamples are
$1.3\times 10^{24}$ and $5.0 \times 10^{24}$~W~Hz$^{-1}$ respectively
(corresponding to total IR luminosities of $2.5\times 10^{10}$ and
$7.9 \times 10^{10}\rm L_{\odot}$), so that they differ by a factor 3 in
typical luminosity. Fig.~\ref{fig:flux250} shows the number counts as
a function of the 250~$\mu$m flux.  Although these two subsets are
well distinguished in luminosity, they have similar distributions of
observed 250~$\mu$m flux.

The separation of the two samples by $\rm L_{250}$ is somewhat
blurred by the uncertainties in the flux measurements and assumed
k-corrections. Perturbing the luminosities according to the
flux measurement errors in the H-ATLAS catalogue \citep{Rigby2010} 
makes little difference with just 5\% of the sample switching
from the bright to the faint subsets. The k-correction depends
on the values of $T$ and $\beta$ assumed in equation~(\ref{eq:Tbeta}).
The sample of \cite{Amblard2010} spans the ranges 
T=28$\pm$8~K and $\beta$=1.4$\pm$0.1. This uncertainty can also scramble
the luminosity subsets somewhat but even the most extreme choice
of $T=36$~K and $\beta=1.5$ only switches 8\% of the sample
from the bright to the faint subsets. We return to the effect
this might have on our clustering results in Section.~\ref{sec:cor}.

\begin{figure}
\bc

\hspace{-0.6cm}
\resizebox{8.5cm}{!}{\includegraphics{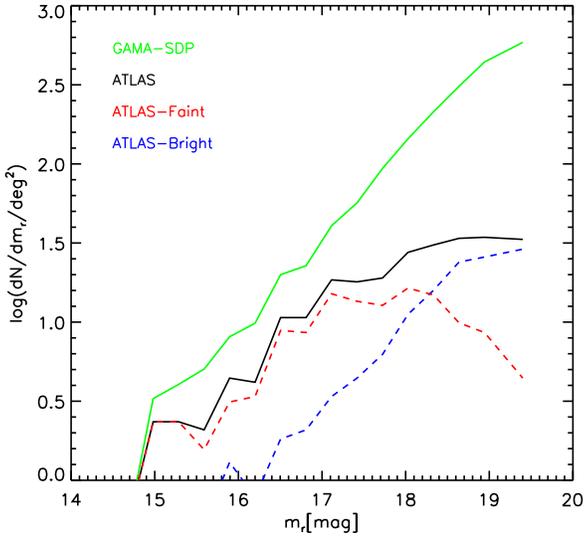}}
\caption{The r-band number counts per magnitude per square degree
versus apparent magnitude.  As in Fig.~\ref{fig:flux250}, the blue and
red dashed curves are for the bright and faint H-ATLAS sources, and
black is the combination of the two. To compare, the green curve gives
the number counts of all GAMA galaxies in the same region.}
\label{fig:rmag}
\ec
\end{figure}

\begin{figure}
\bc
\hspace{-0.6cm}
\resizebox{8.5cm}{!}{\includegraphics{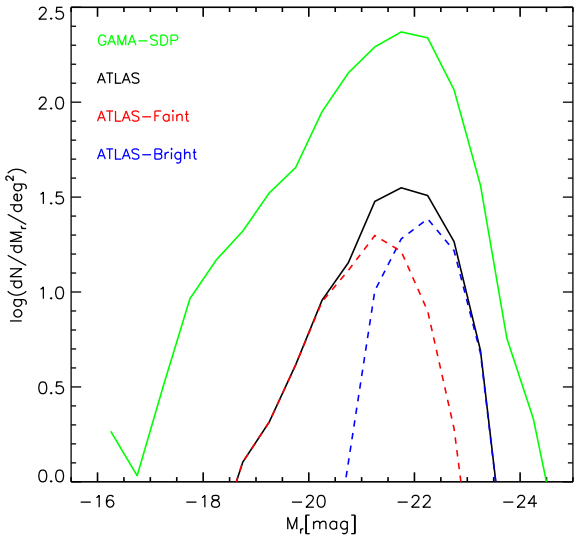}}
\caption{Similar to Fig.~\ref{fig:rmag}, but as a function of r-band absolute
magnitude. The curves are colour coded in the same way as in Fig.~\ref{fig:rmag}. }
\label{fig:Mr}
\ec
\end{figure}

The distributions of apparent and absolute r-band magnitudes
(corrected for Galactic dust extinction) are shown
in Figs.~\ref{fig:rmag} and ~\ref{fig:Mr} respectively. The r-band
absolute magnitudes have been k-corrected to $ z=0$
\citep{Blanton2003}. 
The r-band absolute magnitude for the full H-ATLAS sample peaks around
$-21.7$, somewhat brighter than the Milky Way. For comparison, we also
include the corresponding properties of the full GAMA sample in the
same sky area in Figs.~\ref{fig:rmag} and~\ref{fig:Mr}. It can be seen
that while there are more GAMA than H-ATLAS galaxies, their
distributions of apparent and absolute r-band magnitude are
similar. The extra galaxies in the GAMA catalogue may correspond to
early-type and some late-type galaxies, for which the current SFRs are
very low, leading to their absence from the far-IR survey. More
detailed work on the properties of these galaxies is needed in the
future.

\begin{figure}
\bc
\hspace{-0.6cm}
\resizebox{8.5cm}{!}{\includegraphics{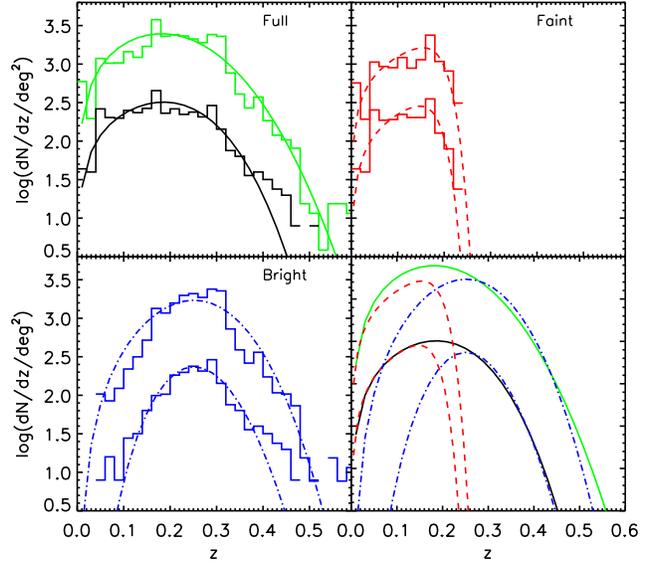}}
\caption{Redshift distributions of each of our samples, plotted as the
number of galaxies per unit redshift per square degree. The colour
coding is as in Fig.~\ref{fig:rmag}: red curves are for the
H-ATLAS-faint and GAMA-faint samples, blue curves are for the
H-ATLAS-bright and GAMA-bright samples, black curves are for the full
H-ATLAS sample and green for the full GAMA sample. For H-ATLAS, the
histograms are the observed distributions, and smooth curves are fits to
these distributions. For GAMA, the histograms show the redshift
distribution of galaxies in the GAMA-SDP field, while smoothed curves
are the fits to the full GAMA area. The fitted curves are collected in
the bottom-right panel.}
\label{fig:Dz}
\ec
\end{figure}

The redshift distributions of our samples are shown in
Fig.~\ref{fig:Dz} as histograms.  In each case, the upper histograms
and curves correspond to the GAMA sample and the lower ones to the
H-ATLAS sample. As expected, the more luminous H-ATLAS galaxies tend
to lie at higher redshifts. The redshift distributions of the luminous
and faint galaxies cross at $z\sim 0.2$, which is roughly the median
value for all the 969 H-ATLAS sources. To help interpret the 
cross-correlation of the faint
and bright H-ATLAS sources with GAMA galaxies, we want subsets of
the GAMA galaxies with similar redshift distributions to the
corresponding H-ATLAS samples. To achieve this, we split the GAMA
sample at M$_{\rm r} = -21.2$\,mag into faint and bright subsets. It
can be seen in Fig.~\ref{fig:Dz} 
that this choice of dividing magnitude results in the
corresponding subsets of H-ATLAS and GAMA samples having very similar
redshift distributions. The full, faint and bright GAMA samples have
median absolute magnitudes M$_{\rm r}$ of $-21.5$, $-20.5$ and
$-22.0$\,mag respectively.

\section{Correlation functions} 
\label{sec:cor}
In this section, we first calculate the auto-correlation functions of
the GAMA and H-ATLAS galaxies, then their cross-correlation, and
finally the clustering bias of the H-ATLAS galaxies. The
auto-correlation of the GAMA galaxies is needed for calculating the
relative bias from the cross-correlation, while the H-ATLAS
auto-correlation provides a consistency check on the results from the
cross-correlation, and also allows us to compare with the
auto-correlation results of \citet{vanKampen2010}.

\subsection{Auto-correlation functions}

In this subsection, we estimate the auto-correlation function of the
GAMA and H-ATLAS SDP samples and, for the H-ATLAS sample, the
dependence of clustering strength on the Herschel 250~$\mu$m
luminosity, $\rm L_{250}$. We begin by considering the correlation
function in redshift space, $\xi(r_{\perp},r_{\parallel})$, where
$r_{\perp}$ and $r_{\parallel}$ are the comoving separations
perpendicular and parallel to the line of sight respectively, and
integrate this over the line-of-sight separation, $r_\parallel$, to
obtain the projected correlation function. This removes the
effect of peculiar velocities on the estimate of the spatial
correlation function. 

There are several estimators for the auto-correlation function in the
literature, all which require the generation of a uniform random
catalogue with the same mask as the galaxy catalogue itself. 
In this work we adopt the estimator proposed by \cite{Hamilton1993}, 

\begin{equation}
\xi(r_{\perp},r_{\parallel}) = \frac{DD(r_{\perp},r_{\parallel})\,
RR(r_{\perp},r_{\parallel})}{[DR(r_{\perp},r_{\parallel})]^2}-1,
\label{eq:wrppi}
\end{equation}
where $DD(r_{\perp},r_{\parallel})$,
$DR(r_{\perp},r_{\parallel})$ and $RR(r_{\perp},r_{\parallel})$ are
counts of data-data, data-random and random-random pairs,
respectively. To generate smooth redshift distributions for the random
samples we fit their redshift distributions with the functional form
\begin{equation}
N(z) \propto  z^{\alpha} \exp(-\beta z ^{\eta}).
\end{equation}
The fits to the redshift distributions of GAMA and H-ATLAS (sub)samples
are shown as smooth curves in Fig.~\ref{fig:Dz}.  To obtain a robust
estimate of the mean redshift distribution of the GAMA galaxies we
made use of the full 144 deg$^2$ of the GAMA catalogue
($\sim$9$\times10^4$ galaxies), rather than just the subset that
overlaps with the H-ATLAS area ($\sim$7$\times10^3$ galaxies). The
completeness mask of \cite{Norberg2010a} was used to generate the random 
catalogue corresponding to the GAMA sample. 

Following standard practice, we estimate the projected correlation
function, $w(r_{\rm p})$, by integrating Eqn.~\ref{eq:wrppi} along line
of sight $r_{\parallel}$:
\begin{equation}
w(r_{\rm p}) =
w(r_\perp)= \int^{\infty}_{-\infty}\xi(r_{\perp},r_{\parallel}) \,
{\rm d}r_{\parallel} .
\label{eq:integrate}
\end{equation}
In reality, we cannot integrate to infinity. Instead, we have chosen
to integrate to $\pm$50~Mpc, but we test the impact of varying this
limit.  Errors are estimated using the Jackknife technique. We split
each galaxy sample into 16 equal area regions and then calculate the
correlation functions for data taken from any 15 of these 16
regions. The scaled scatter of the Jackknife samples gives an estimate
of the errors on the corresponding correlation functions
\citep[e.g.][]{Norberg2009}.

The projected correlation function is  related to the real-space
correlation function by a simple Abel transform (Peebles 1980).  For a
power law, $w(r_p)=Ar_p^{1-\gamma}$, the 3-D correlation function, 
$\xi(r)$, is also a power law, $\xi(r) = (r/r_0)^{-\gamma}$. The
parameters are related by
\begin{equation}
r_0^{\gamma} = \frac{A\Gamma(\gamma/2)}{\Gamma(1/2)\Gamma[(\gamma-1)/2]},
\label{eq:3dcor}
\end{equation}
where $\Gamma(x)$ is the standard Gamma function. 

\begin{figure*}
\bc
\hspace{-0.6cm}
\resizebox{14.5cm}{!}{\includegraphics{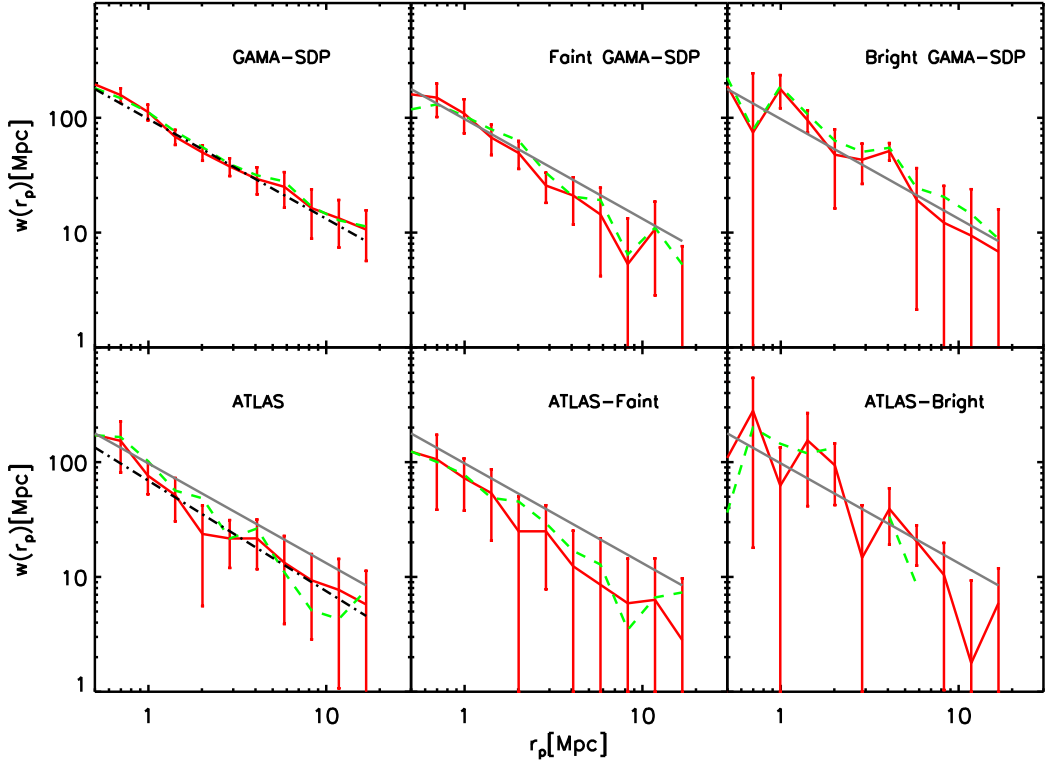}}
\caption{Two-point projected auto-correlation functions. From left to
  right and top to bottom, they correspond to the GAMA, GAMA-faint and
  GAMA-bright; H-ATLAS, H-ATLAS-faint and H-ATLAS-bright samples
  respectively. The red curves show the result of truncating the
  line-of-sight integration in Eqn.~4 at 50~Mpc and the green dashed
  curves at 100~Mpc. The black dash-dotted lines are power-law fits to
  the data in red for full GAMA and H-ATLAS samples only. To aid
  comparison, the fit to the GAMA auto-correlation function for the
  full sample (top left panel) is repeated as a grey curve in the other panels. Error
  bars are estimated using the Jackknife technique. }
\label{fig:corauto}
\ec
\end{figure*}

The two-point projected auto-correlation functions are plotted in
Fig.~\ref{fig:corauto} as red curves. To test the convergence of the
line-of-sight integral in Eqn.~\ref{eq:integrate}, we show with green
curves (here and later also in Fig.~\ref{fig:corcross}) the result of
extending the integration out to 100~Mpc. The projected correlation
function is seen to be insensitive to the precise choice of
integration limit. 

For the GAMA-SDP sample (top row), the projected correlation functions
are measured in the region of overlap with H-ATLAS. The reason for
remeasuring the GAMA correlation functions in this restricted area
rather than showing the less noisy estimate from the full GAMA dataset
\citep{Norberg2010a} is that we are interested in the relative clustering
of H-ATLAS and GAMA galaxies and this choice will reduce the impact of
sample variance on the comparison. The GAMA-SDP correlation function
can be well fitted with a power law (black dot-dashed curve). Fitting
Eqn.~\ref{eq:integrate} to the data in the range (1 -- 12)\,~Mpc and
using Eqn.~\ref{eq:3dcor}, we find $r_0$ = 5.96 $\pm$ 0.62 Mpc and
$\gamma$ = 1.87 $\pm$ 0.21. This is consistent with the values for
$L^*$ optical galaxies (with M$_{\rm r}^{0.1} \approx -21.1$)
estimated in the SDSS: $r_0 = 6.6 \pm 0.3$~Mpc and $\gamma = 1.87 \pm
0.03$ \citep{Zehavi2010}. 


The lower panels of Fig.~\ref{fig:corauto} show the projected
auto-correlation functions for our three H-ATLAS samples. The best-fit
values of $r_0$ and $\gamma$ for these are summarised in
Table~1. \citet{vanKampen2010} have carried out a more detailed
analysis of the auto-correlation function of H-ATLAS galaxies using
the angular correlations in redshift slices.
They obtain a best estimate of the spatial clustering length, averaged
over the redshift range $0.1<z<0.3$, of $r_0 = 5.5 \pm 0.9$~Mpc. Our
estimate of $r_0$ given in Table~1 for the auto-correlation of our
full H-ATLAS sample is consistent with this.

For comparison, the best-fit power law for the GAMA-SDP
auto-correlation function for the full GAMA sample is reproduced by a
grey line in all of the other panels of Fig.~\ref{fig:corauto}. The
full H-ATLAS sample is somewhat less clustered than the full GAMA-SDP
sample. The faint H-ATLAS galaxies appear to have similar clustering
to the full H-ATLAS sample, while the bright H-ATLAS galaxies appear
to be more strongly clustered. However, the statistical uncertainties
in the estimates for these small samples are clearly rather large and,
moreover, systematic errors could be introduced by fitting smooth
curves to their noisy redshift distributions. These limitations are
largely overcome in the next section where we measure the clustering
of the H-ATLAS galaxies by cross-correlating with the much larger
GAMA-SDP sample. Furthermore, by estimating the GAMA-SDP radial selection function using the full GAMA survey covering an area about 10 times larger than the GAMA-SDP region, systematic uncertainties in the modelling of the radial selection function are significantly reduced.

In Fig 6 it is apparent that our jackknife error bars are sometimes
noisy as witnessed e.g. by the large error bars at$\sim$0.7Mpc or
$>$6Mpc for the GAMA bright sample, or by the small error bars on the
GAMA faint sample on scales below 2Mpc. Further investigation has
revealed that this is a result of our small sample and occurs because
the clustering on particular scales can be dominated by one or two
structures and so vary significantly in just one or two of our
jackknife samples. Such fluctuations are smaller for our
cross-correlation samples, discussed below. Thus the errors quoted for the correlation
length, $r_0$, for the bright and faint GAMA auto-correlation samples
have significant uncertainty, but the cross-correlation results and
their error bars are more robust. The diagnostic tests used for the
robustness of the clustering errors are similar to those presented in
\cite{Norberg2010a}.

\subsection{Cross-correlation functions} 

The cross-correlation function in redshift space of H-ATLAS with GAMA
galaxies is estimated using
\begin{equation}
\xi(r_{\perp},r_{\parallel}) =
\frac{HG(r_{\perp},r_{\parallel})\,RR(r_{\perp},r_{\parallel})}{HR(r_{\perp},r_{\parallel})\,
GR(r_{\perp},r_{\parallel})}-1 , 
\end{equation}
where HG, HR, GR, RR are counts of H-ATLAS-GAMA, H-ATLAS-random,
GAMA-random and random-random pairs, respectively. In each case, the
random sample is generated so as to match the redshift distribution of
the GAMA galaxies. Thus, for our estimates of the cross-correlation
functions, at no point do we need to fit the noisy redshift
distributions of the small samples of H-ATLAS galaxies.  As for the
auto-correlation functions, we calculate the projected two-point
cross-correlation functions according to Eqn.~\ref{eq:integrate} and
estimate the errors using the jackknife technique.

\begin{figure}
\bc
\hspace{+0.6cm}
\resizebox{8.5cm}{!}{\includegraphics{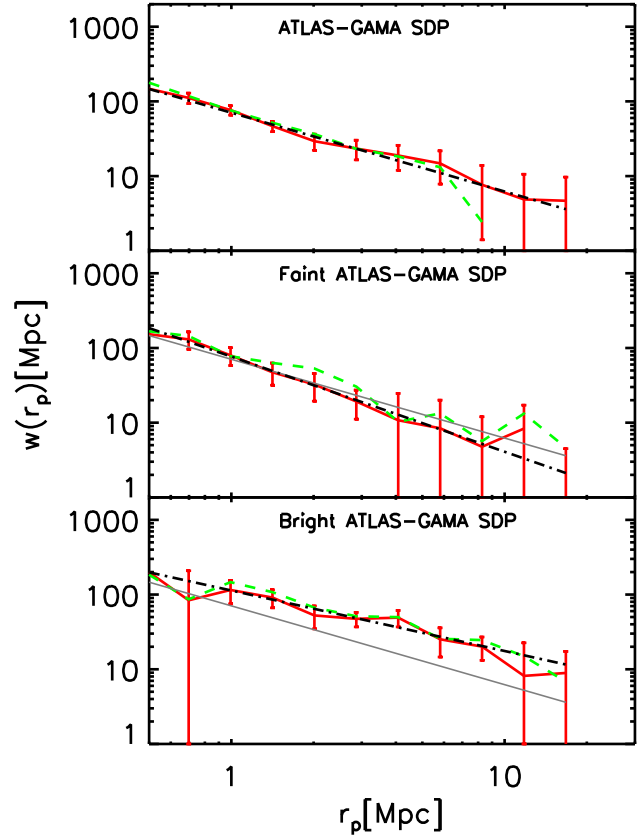}}
\caption{Two-point projected cross-correlation of all H-ATLAS with all
  GAMA galaxies (top), faint H-ATLAS with faint GAMA galaxies (middle)
  and bright H-ATLAS with bright GAMA galaxies (bottom). As in
  Fig.~\ref{fig:corauto}, red curves show the result of intergrating
  Eqn.~\ref{eq:integrate} to 50~Mpc and the green dashed curves to 100~Mpc. Black
  dash-dotted curves show the power-law fits to the red curves over
  the range 1-12~Mpc. For comparison, the fit to the all H-ATLAS-GAMA
  cross-correlation function is replicated as grey lines in the lower
  two panels. Error bars are estimated using the Jackknife technique.}
\label{fig:corcross}
\ec
\end{figure}

The projected cross-correlation functions are shown in
Fig.~\ref{fig:corcross}. The top panel shows the GAMA-H-ATLAS result
when the limit of integration in Eqn.~\ref{eq:integrate} is taken to
be 50~Mpc (red curves) and 100~Mpc (green curves) respectively. The
dot-dashed line is the best fitting power law to the 50~Mpc
estimate. It shows that the H-ATLAS-GAMA cross-correlation function is
well fitted by a power law, with $r_0$ = 4.63~$\pm$~0.51 Mpc and
$\gamma$ = 2.05~$\pm$~0.31, indicating that the clustering of the
H-ATLAS galaxies is weaker than that of GAMA-SDP galaxies. This
inferred difference between the strength of the H-ATLAS and GAMA-SDP
clustering appears larger than suggested by comparing the upper and
lower left panels of Fig.~\ref{fig:corauto}, or the values of $r_0$ in
Table~1. This might be the result
of a bias in the redshift distribution of the random samples for the
H-ATLAS galaxies, which is obtained by fitting a smooth function to
noisy data (\S~3.1).

The lower panels in Fig.~\ref{fig:corcross} show cross-correlation
functions for subsets of luminous and faint H-ATLAS and GAMA
galaxies. For comparison, this best-fit line to the GAMA-H-ATLAS
function is replicated in grey in these panels.  Again, we find that
the clustering of faint H-ATLAS galaxies is weaker than that of the
bright galaxies. The estimates of $r_0$ and $\gamma$ for these samples
are summarized in Table~1.

As discussed in Section~\ref{sec:sample}, there are uncertainties in
  the $250~\mu$m k-correction and the
  flux measurements. Adopting the most extreme perturbation to the
  k-corrections ($T=36$~K and $\beta=1.5$, see
  Section~\ref{sec:sample}) and perturbing the fluxes according to
  the measurement errors quoted in \cite{Rigby2010} 
  shifts the $r_0$ values of our estimated H-ATLAS autocorrelation
  functions by an amount comparable to the quoted 1-$\sigma$ statistical
  uncertainty. This variation is largely caused by the limited 
  size of these samples and the resulting uncertainty in fitting their redshift
  distrbutions. The cross-correlations on which we focus, and which do
  not depend on the redshift distributions of the H-ATLAS samples,
  are much
  less affected by the uncertainties in the k-corrections and
  flux measurements. In this case, the same perturbations affect the 
  $r_0$ values, by no more than 15\% of their quoted statistical error
  and so make a negligible contribution to the uncertainty in our results.

\subsection{Bias of H-ATLAS galaxies}

To interpret the meaning of the estimated large-scale
cross-correlation functions, consider the simple linear bias model in
which the auto- and cross-correlation functions of H-ATLAS and GAMA
galaxies are related to the auto-correlation function, $\xi_{\rm m}$, 
of the mass at redshift $z=0$ by
\begin{eqnarray}
\xi_{\rm H}({\bf r}) &=& b_{\rm H}(z)^2 \, D^2(z) \, \xi_{\rm m} ({\bf
r})  \\
\xi_{\rm G}({\bf r}) &=& b_{\rm G}(z)^2 \, D^2(z) \, \xi_{\rm m} ({\bf
r})  \\
\xi_{\rm HG}({\bf r}) &=& b_{\rm H}(z) b_{\rm G}(z) \, D^2(z) \, \xi_{\rm m} ({\bf r}) ,
\end{eqnarray}
where the subscripts ${\rm H}$ and ${\rm G}$ denote H-ATLAS or GAMA
respectively, $D(z)$ is the linear growth factor of the perturbations
in the mass, and $\xi_{\rm m} ({\bf r})$ is the auto-correlation
function of the dark matter. In this case, the projected
cross-correlation function that we have estimated is related to that
of the mass at $z=0$ through
\begin{equation}
 w_{{\rm HG}}(r_{\rm p}) = \langle b_{\rm H} b_{\rm G} D^2(z)
\rangle w_{\rm
 m}(r_{\rm p}),
\end{equation}
where the average product of the bias and growth factors is given by  
\begin{equation} 
\label{eq:bias}
\langle b_{\rm G}  b_{\rm H}  D^2(z) \rangle =
\frac{
 \int \bar{n}_{\rm G} (z) \bar{n}_{\rm H} (z) b_{\rm G} (z) b_{\rm H}
 (z) D^2(z)
\left(\frac{{\rm d}V}{{\rm d}z} \right) {\rm d}z
}{\int  \bar{n}_{\rm G} (z) \bar{n}_{\rm H} (z) \left(\frac{{\rm
d}V}{{\rm d}z} \right) {\rm d}z}, 
\end{equation}
and $\bar{n}_{\rm H} (z)$ and $\bar{n}_{\rm G} (z)$ are the mean space
densities of the H-ATLAS and GAMA samples at redshift $z$.
For the auto-correlation function of the GAMA galaxies this reduces
to
\begin{equation}
 w_{{\rm G}}(r_{\rm p}) = \langle b^2_{\rm G} D^2(z)
\rangle w_{\rm
 m}(r_{\rm p}),
\end{equation}
where 
\begin{equation} 
\langle b^2_{\rm G}  D^2(z) \rangle =
\frac{
 \int \bar{n}^2_{\rm G} (z)  b^2_{\rm G} (z) D^2(z)
\left(\frac{{\rm d}V}{{\rm d}z} \right) {\rm d}z
}{\int \bar{n}^2_{\rm G} (z) \left(\frac{{\rm d}V}{{\rm d}z} \right) {\rm d}z}.
\end{equation}

The relative bias of the H-ATLAS and GAMA galaxies is then, 
\begin{equation}
b^{\rm rel}_{\rm HG}
= w_{\rm GH}(r_p) /w_{\rm GG}(r_p) =
\langle b_{\rm H}   b_{\rm G} D^2(z) \rangle /\langle b_{\rm G}^2 D^2(z)
\rangle .
\end{equation}
In principle, this depends on both the bias parameters $b_{\rm H}$,
$b_{\rm G}$ and on $D(z)$. However, since by construction the redshift
distributions of the full/faint/bright H-ATLAS samples match well with
those of the corresponding (full/faint/bright) GAMA samples, the
dependence on $D(z)$ will approximately cancel. If the bias parameters
$b_{\rm H}$ and $b_{\rm G}$ evolve with redshift in the same way, then
this evolution will also approximately cancel out in the relative
bias. This is the reason why we cross correlate H-ATLAS faint/bright
with GAMA faint/bright instead of all GAMA galaxies. 

We estimate the mean relative bias $\bar{b}^{\rm rel}_{\rm
HG}$ of H-ATLAS and GAMA galaxies using, 
\begin{equation} 
  \bar{b}^{\rm rel}_{\rm HG} = \frac{\Sigma b^{\rm rel,i}_{\rm HG}
/\sigma_{\rm i}^2}{\Sigma 1/\sigma_{\rm i}^2}, 
\label{eq:brel}
\end{equation}
where $b^{\rm rel,i}_{\rm HG}$ is obtained directly from the measured
projected H-ATLAS-GAMA cross-correlation function and the GAMA
auto-correlation function (rather than from the fits given in
Table~1), and $\sigma_{\rm i}$ represents the Jackknife error on
$b^{\rm rel,i}_{\rm HG}$ estimated at each pair separation. This simple estimator ignores 
correlations between the measurements at different separations and so may not be optimal, but we do take account of such correlations in estimating the error on $\bar{b}^{\rm rel}_{\rm HG}$. Our error
on $\bar{b}^{\rm rel}_{\rm HG}$ is estimated using the Jackknife
technique, by calculating the mean $b^{\rm rel}_{{\rm HG},j}$ for each
Jackknife sample (assuming the same values of $\sigma_{\rm i}$ as used
in Eqn.~\ref{eq:brel}) and then looking at the scatter in values
between Jackknife samples. 

\begin{figure}
\bc
\hspace{-0.6cm}
\resizebox{8.5cm}{!}{\includegraphics{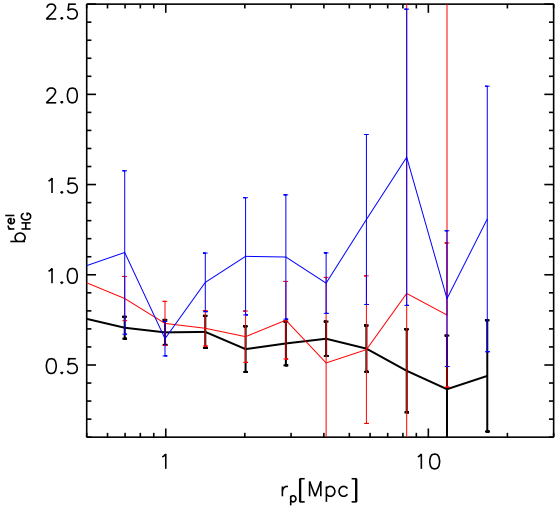}}
\caption{The relative bias $b^{\rm rel}_{\rm HG}$ estimated from the
ratio of the projected H-ATLAS-GAMA cross-correlation function,
$w_{\rm HG}(r_{\rm p})$, to the corresponding GAMA auto-correlation
function $w_{\rm G}(r_{\rm p})$.  The black curve is for the bias of
the full H-ATLAS sample relative to the full GAMA sample, while the
blue curve is for the bright H-ATLAS galaxies relative to bright GAMA
galaxies, and the red curve is for faint H-ATLAS galaxies relative to
faint GAMA galaxies.  Errors are estimated using the Jackknife
technique.}
\label{fig:bias}
\ec
\end{figure}

Our estimates of $\bar{b}^{\rm rel}_{\rm HG}$ are shown in
Fig.~\ref{fig:bias}. For the full H-ATLAS and GAMA samples, the mean
relative bias over the range of separations 2-8~Mpc, where the
two-halo term dominates and where we have good statistics, is $b^{\rm
rel}_{\rm HG}(all) = 0.61\pm 0.08$. Thus, we conclude that the
clustering strength of H-ATLAS galaxies is significantly weaker than
that of GAMA SDP galaxies. This important conclusion is revealed only
by taking advantage of the cross-correlation function technique. As
shown in Table~1, our estimates of the auto-correlation functions are
much too noisy (and probably subject to systematic errors) to detect any
difference between the two galaxy samples. From the cross-correlation
of the faint H-ATLAS with the faint GAMA samples, we obtain a relative
bias of $b^{\rm rel}_{\rm HG}(faint) = 0.67\pm 0.13$, while from the
cross-correlation of the bright H-ATLAS with the bright GAMA galaxies,
we obtain $b^{\rm rel}_{\rm HG}(bright) = 1.04\pm 0.22$.


To convert the estimates of relative bias into values of the absolute
bias for the different H-ATLAS samples, we need to know the absolute
bias of the different GAMA samples. For this we use the results of
\citet{Zehavi2010}, who measured the clustering as a function of
r-band luminosity in the SDSS, and combined that with a theoretical
prediction for the clustering of the dark matter in the $\Lambda$CDM
cosmology. An important qualification is that the values of bias
measured by \citeauthor{Zehavi2010} effectively apply at the average
redshift of the SDSS, $z\sim 0.1$. The bias of r-band selected
galaxies is expected to evolve with redshift, but quantifying the size
of this effect for the redshift range $z\lsim 0.5$ probed in the
present paper must await a detailed clustering analysis of the full
GAMA redshift survey. Here, we will simply assume that the bias
factors for GAMA and H-ATLAS galaxies can be taken to be constant over
the redshift range studied here. We therefore use Eqn.~(10) from
\citeauthor{Zehavi2010}, scaled to $\sigma_8=0.8$, to calculate the
value of the bias as a function of r-band absolute magnitude.

Our full, faint and bright GAMA samples have median absolute
magnitudes M$_{\rm r}^{0.1}=-21.3$, $-20.3$ and $-21.8$ respectively
(k-corrected to $z=0.1$ to be consistent with
\citeauthor{Zehavi2010}), implying average r-band bias factors of
$b_{\rm G}=1.17$, $1.05$ and $1.29$. This then leads to absolute bias
values of $b_{\rm H}= 0.71 \pm 0.09$, $0.70 \pm 0.14$ and $1.34 \pm
0.28$ respectively for the full, faint and bright H-ATLAS
subsamples. We find that the bright H-ATLAS galaxies are more strongly
clustered than the H-ATLAS population as a whole at the 2$\sigma$
level, which confirms the trend seen from the H-ATLAS auto-correlation
functions in Fig.~\ref{fig:corauto}. This result implies that the
excess clustering of the bright H-ATLAS galaxies reflects a genuine
and strong dependence of clustering on {\em far-infrared luminosity}
and thus on star formation rate. We detect no significant difference
between the bias of the faint H-ATLAS galaxies and that of the
population as a whole. This result, however, could be affected by our
assumption of a constant bias over the redshift of interest.


The final step is to use the estimated clustering bias of H-ATLAS
galaxies to constrain the masses of the halos hosting them. In the
$\Lambda$CDM model at the present day, the halo bias is a very weak function of halo mass
for halos less massive than $10^{12}$M$_{\odot}$ and increases rapidly
with increasing halo mass at higher masses \citep{Mo_White2002}. Using
the fitting formula for bias as a function of halo mass at $z=0$ from
\citet{Seljak2004}, obtained from simulations of a $\Lambda$CDM
universe, we infer an average host halo mass $\log_{10} M/{\rm M_\odot}
\approx 12.1^{+0.5}_{-\infty}$ (or a 2$\sigma$ upper limit $\log_{10}
M/{\rm M_\odot} \lsim 12.8$) for the full H-ATLAS sample. We find very
similar values for the faint H-ATLAS subsample, $\log_{10} M/{\rm M_\odot}
\approx 12.0^{+0.71}_{-\infty}$ (or a 2$\sigma$ upper limit $\log_{10}
M/{\rm M_\odot} \lsim 13.0$).  For the bright H-ATLAS sample, the
average halo mass is $\log_{10} M/{\rm M_\odot} \approx
13.6^{+0.3}_{-0.4}$. The more luminous H-ATLAS galaxies thus appear to
be hosted in significantly more massive halos than the faint ones.
 Note that given the large errors in the estimates of halo masses, it is reasonable that the 2$\sigma$ upper limits on the host masses of the faint and bright subsamples are both higher than that of the full sample.


\begin{table*}
\caption{The correlation length, $r_0$, and slope $\gamma$, 
for the power-law fits 
to our auto- and cross-correlation functions and the mean redshift,
$z_{\rm mean}$, number of galaxies, $N_{\rm gal}$, and relative bias of each sample.} 
\begin{tabular}{||l||c||c||c||c||c||} 
\hline
 CF & $r_0$[Mpc] &$\gamma$ & $z_{\rm mean}$ & $N_{\rm gal}$ & relative bias \\
\hline

 GAMA-SDP auto & 5.96 $\pm$ 0.62 & 1.87 $\pm$ 0.21 & 0.21 & 7761 &\\
 H-ATLAS auto & 4.76 $\pm$ 0.63 & 1.96 $\pm$ 0.38  & 0.19 & 970& \\
 Faint GAMA-SDP auto & 5.19 $\pm$ 0.77 & 2.20  $\pm$ 0.43 & 0.13  & 1981 &\\
 Bright GAMA-SDP auto & 7.06 $\pm$ 0.45 & 1.90 $\pm$ 0.27  & 0.26 & 4780 &\\
 Faint H-ATLAS auto & 4.49 $\pm$ 1.05 & 2.15 $\pm$ 0.54  & 0.12  & 484 &\\
 Bright H-ATLAS auto & 5.72 $\pm$ 0.53 & 2.06 $\pm$ 0.27 & 0.26  & 485 &\\  \\ 

 H-ATLAS-GAMA cross & 4.63 $\pm$ 0.51 & 2.05 $\pm$ 0.31 & &  & 0.61 $\pm$ 0.08 \\
 Faint H-ATLAS - Faint GAMA cross  & 4.38  $\pm$ 0.77 & 2.27 $\pm$ 0.47 && & 0.67 $\pm$ 0.13\\
 Bright H-ATLAS - Bright GAMA cross  & 6.68 $\pm$ 0.44 & 1.81 $\pm$ 0.26 && &1.04 $\pm$ 0.22\\

\hline
\end{tabular} 
\label{table:para}
\end{table*}

\section{Conclusions}
We have used a subset of the H-ATLAS galaxies in the SDP field, which
have spectroscopic redshifts from the optical GAMA redshift survey, to
calculate the projected cross-correlation functions of far-IR and
optically selected galaxies.  We find that these H-ATLAS galaxies
(which have a median redshift $z\approx 0.2$, median 250~$\mu$m
luminosity $\rm L_{250} \approx 2.5 \times 10^{24}$~W~Hz$^{-1}$,
and median total IR luminosity $\rm L_{IR} \sim 5.0 \times 10^{10}\rm
L_{\odot}$) are significantly less strongly clustered than the
optically selected GAMA galaxies (which have a median absolute
magnitude, M$_{\rm r}=-21.5$\,mag) at the same redshifts. This effect
is also seen (though with lower significance) in the auto-correlations
of the H-ATLAS and GAMA galaxies.

From the cross-correlation analysis, combined with the previously
measured clustering of optical galaxies in the SDSS, we find that
H-ATLAS galaxies are less clustered than the dark matter, with an
average bias $b = 0.71\pm 0.09$. This implies a typical host halo mass of $\sim 1.25 \times 10^{12}\rm M_{\odot}$ for the
H-ATLAS galaxies in our sample (which are mostly at low redshift),
comparable to the halo of the Milky Way. These preliminary results for
the host halo masses of the H-ATLAS galaxies are consistent with the
theoretical predictions of \citet{lacey2010}, who find a typical halo
mass of $1.6 \times 10^{12}{\rm M}_{\odot}$.  (Note that \citet{lacey2010} used the
halo bias formula of \cite{Sheth2001} which predicts a somewhat larger
bias than the \cite{Seljak2004} formula used here at low masses.)


We also split our H-ATLAS sample into subsamples of high and low
far-IR luminosity, and investigate their clustering properties.  Both
the cross- and auto-correlation functions suggest a dependence of
clustering on far-IR luminosity over the range $\rm
L_{IR}=2.5 \times 10^{10}-7.9 \times 10^{10}\rm L_{\odot}$, with the bright galaxies
being more strongly clustered than the faint ones at 2$\sigma$
significance, implying that the more luminous galaxies are hosted by
more massive dark halos. The average halo mass for the bright sample
is around 4 $\times $10$^{13}$M$_{\odot}$ and the 2$\sigma$ upper limit for the
halos hosting the faint sample is $10^{13}{\rm M}_{\odot}$.
The dependence
of clustering on far-IR luminosity that we find here appears
significantly stronger than the model predictions of \cite{lacey2010}
who find $M_{\rm halo} \sim 1.3 \times 10^{12}$ and $2.0 \times 10^{12} {\rm M}_{\odot}$ for
galaxies of comparable luminosities to our faint and bright
subsamples. It will be interesting to test whether this discrepancy
persists in the full H-ATLAS survey. 
 As luminosity and redshift are correlated in a flux limited
 sample, our high L$_{250}$ luminosity subset has a higher median
redshift than its fainter counterpart. Hence, in principle,
strong evolution of clustering with redshift could be contributing
to our inferred dependence of clustering on luminosity.
We will be able to directly address this ambiguity with the much larger full H-ATLAS
sample by splitting the sample into redshift bins.
When completed, this survey will
enable comprehensive investigations of the clustering and environments
of star-forming galaxies.

\section*{Acknowledgments}
CSF acknowledges a Royal Society Wolfson Research Merit award.  SC
acknowledges the support of the Leverhulme Research Fellowship.
This
work was supported in part by a rolling grant from the Science and
Technology Facilities Council to the ICC. The Herschel-ATLAS
survey\footnote{http://www.h-atlas.org/} is being carried with ESA's
Herschel Space Observatory, which is equipped with instruments
provided by European-led Principal Investigator consortia, with
important participation from NASA. U.S.A. authors acknowledge support
provided by NASA through JPL.
GAMA\footnote{http://www.gama-survey.org/} is a joint
European-Australian spectroscopic campaign using the Anglo-Australian
Telescope. The GAMA input catalogue is based on data from the Sloan
Digital Sky Survey and the UKIRT Infrared Deep Sky
Survey. Complementary imaging of the GAMA regions is being obtained by
a number of independent surveys including GALEX MIS, VST KIDS, VISTA
VIKING, WISE, Herschel ATLAS, GMRT and ASKAP, providing UV to radio
coverage. PN acknoweldge the support of a Royal Society University Research Fellowship. We thank Douglas Scott for critical reading our manuscript. 
\bibliographystyle{mn2e}

\setlength{\bibhang}{2.0em}
\setlength\labelwidth{0.0em}

\bibliography{cor_cross}

\end{document}